\documentclass[preprint]{aastex}
\usepackage{amssymb,amsmath}
\usepackage{graphicx}
\usepackage{natbib}
\usepackage[OT1,T1]{fontenc}

\begin{document}

\title{Bidirectional outflows as evidence of magnetic reconnection leading to a solar microflare}
\author{Jie Hong\altaffilmark{1,2}, M.~D. Ding\altaffilmark{1,2}, Ying Li\altaffilmark{1,2}, Kai Yang\altaffilmark{1,2}, Xin Cheng\altaffilmark{1,2}, Feng Chen\altaffilmark{3}, Cheng Fang\altaffilmark{1,2}, and Wenda Cao\altaffilmark{4}}
\affil{\altaffilmark{1}School of Astronomy and Space Science, Nanjing University, Nanjing 210023, China \email{dmd@nju.edu.cn}}
\affil{\altaffilmark{2}Key Laboratory for Modern Astronomy and Astrophysics (Nanjing University), Ministry of Education, Nanjing 210023, China}
\affil{\altaffilmark{3}Max-Plank-Institut f\"{u}r Sonnensystemforschung, 37077, G\"{o}ttingen, Germany}
\affil{\altaffilmark{4}Big Bear Solar Observatory, New Jersey Institute of Technology, 40386 North Shore Lane, Big Bear City, CA 92314-9672, USA}

\begin{abstract}
    Magnetic reconnection is a rapid energy release process that is believed to be responsible for flares on the Sun and stars. Nevertheless, such flare-related reconnection is mostly detected to occur in the corona, while there have been few studies concerning the reconnection in the chromosphere or photosphere. Here we present both spectroscopic and imaging observations of magnetic reconnection in the chromosphere leading to a microflare. During the flare peak time, chromospheric line profiles show significant blueshifted/redshifted components on the two sides of the flaring site, corresponding to upflows and downflows with velocities of $\pm$(70--80) km s$^{-1}$, comparable with the local Alfv\'{e}n speed as expected by the reconnection in the chromosphere. The three-dimensional nonlinear force-free field configuration further discloses twisted field lines (a flux rope) at a low altitude, cospatial with the dark threads in \ion{He}{1} 10830 \r{A} images. The instability of the flux rope may initiate the flare-related reconnection. These observations provide clear evidence of magnetic reconnection in the chromosphere and show the similar mechanisms of a microflare to those of major flares.
\end{abstract}


\section{Introduction}
Magnetic reconnection is a fundamental physical process on the Sun, and is considered to be the cause of many kinds of solar activities, from drastic eruptions like flares and coronal mass ejections \citep{1964carmichael,1966sturrock,1974hirayama,1976kopp,1999shibata}, to small-scale events like Ellerman bombs \citep{2009archontis,2006fanga} and microflares \citep{1988tandberg}. Although magnetic reconnection is supposed to be the most important ingredient in theoretical models for various solar activities, in particular solar flares, its observational evidence is still elusive. Most major solar flares are believed to originate in the tenuous and fully ionized corona, a favorable place for magnetic reconnection to proceed. There have been some observations supporting the reconnection picture like double hard X-ray sources above the flare loop top \citep{1992sakao,1994masuda,1994bentley}, cusp-like structure in soft X-ray loops \citep{1992acton,1992tsuneta,1995doschek}, reconnection inflows \citep{2001yokoyama,2015sun}, and sudden changes in magnetic topology \citep{1994masuda,2015sun,2011cheng}. Recent observations reveal hot outflows with large velocities during flares, which can be seen in both images and spectra \citep{2007wang,2013liu,2015reeves}. Note that, the outflows in flare-related reconnection have been detected mostly in one direction at a time. In particular, simultaneous bidirectional outflows have been reported in transition region explosive events from an optically thin line, e.g., the \ion{Si}{4} 1393 \r{A} line \citep{1997innes}.

It is known that, theoretically, magnetic reconnection can also occur in the lower atmosphere \citep{1999sturrock}. The most recent observations of Ellerman bombs and hot explosions show bidirectional flows as revealed in chromospheric and coronal lines. These flows indicate magnetic reconnection in the photosphere or the temperature minimum region, which is locally heated to a high temperature comparable with that of the upper chromosphere or transition region \citep{2015vissers,2014peter}. Nevertheless, we still lack enough observational evidence of magnetic reconnection in the lower atmosphere leading to flares.

In this Letter, we report observations of a reconnection scenario in the chromosphere that produces a microflare by investigating the H$\alpha$ and \ion{Ca}{2} 8542 \r{A} spectra. In Section~2, we present the observations. Data analysis and results are shown in Section~3, which is followed by a summary and discussions in Section~4.

\section{Observations}
A microflare of GOES B-class occurred in Active Region NOAA 12146 at 21:04 UT on 2014 August 24. This active region was very complex and had produced three C-class flares during the past three days. We observed this microflare with the Fast Imaging Solar Spectrograph (FISS, \citealt{2013chae}) and the Near InfraRed Imaging Spectropolarimeter (NIRIS, \citealt{2012cao}) installed at the 1.6 meter New Solar Telescope (NST, \citealt{2010cao,2012goode}) of Big Bear Solar Observatory. FISS adopts an Echelle disperser, and can acquire both H$\alpha$ and \ion{Ca}{2}~8542~\r{A} line spectra along a slit simultaneously with the aid of dual cameras. The scan over the target region yields two-dimensional spectra and monochromatic images at different wavelengths. The spectral resolution is 0.19 m\r{A} and 0.26 m\r{A} for the two lines respectively, and the spatial sampling is $0\arcsec.16$. NIRIS can provide \ion{He}{1} 10830 \r{A} images with a high spatial resolution. It covers a field of view (FOV) of about $85\arcsec\times85\arcsec$ with a pixel size of about $0\arcsec.083$. The time cadence is 35 s for FISS and 10 s for NIRIS.

The Atmospheric Imaging Assembly (AIA, \citealt{2012lemen}) on the Solar Dynamics Observatory (SDO, \citealt{2012pesnell}) can provide extreme ultraviolet (EUV) images with a pixel size of $0\arcsec.6$ and a cadence of 12 s, and the Helioseismic and Magnetic Imager (HMI, \citealt{2012schou}) provides both line-of-sight magnetograms and vector field data in the photosphere with the same pixel size but a cadence of 45 s and 12 min respectively.

Before data analysis, we need to co-align the images from different instruments. Note that for the NST data, a de-rotation of images should be performed first. The basic method used for co-alignment is to calculate the correlation coefficients of two images with different offsets, and to find the optimal offset value that corresponds to the maximum coefficient. A sub-region of images containing the most distinguishable feature like a sunspot is used for co-alignment. In practice, we choose a base (reference) image and co-align all the other images with that. As the morphology of the sunspot does not change much during our observations, this code can yield good results. For co-alignment of data from NST and SDO , we use the AIA 1700 \r{A} image as the reference one. The accuracy of all the co-alignment is within $1\arcsec.2$.

\begin{figure}
    \centering
    \epsscale{0.8}
    \plotone{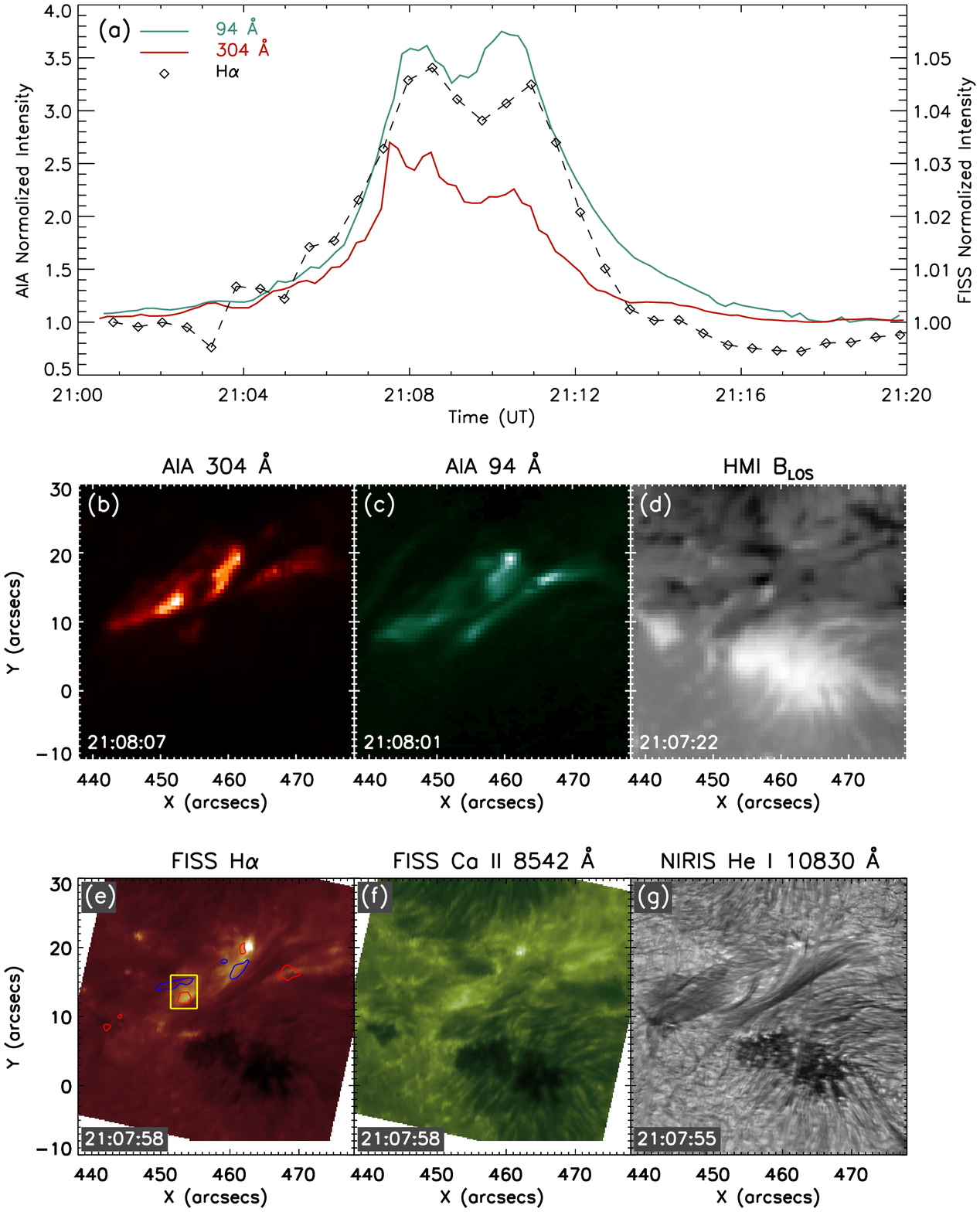}
    \caption{(a) Intensity integrated over the flare region (the FOV in (b)--(g)) as a function of time. Plotted are the intensities at 94 \AA, 304 \AA\ and H$\alpha$, normalized to their corresponding values at the pre-flare time. (b)--(d) The AIA observations in two EUV channels and the line-of-sight magnetogram at the flare peak time. (e)--(f) The H$\alpha$ and \ion{Ca}{2} 8542 \r{A} images reconstructed from the FISS observations at the flare peak time. Also shown in panel (e) are the velocity contour levels of $\pm$10 km s$^{-1}$ derived from the bisector method (red for downflows and blue for upflows). The small yellow box refers to the ROI for a further study. (g) The \ion{He}{1} 10830 \r{A} image from the NIRIS observations at the flare peak time.}\label{fig1}
\end{figure}

Fig.~\ref{fig1}(a) shows the lightcurves of the microflare in two channels of AIA, as well as the lightcurve of H$\alpha$ (wavelength-integrated intensity). It is seen that the microflare lasts for about ten minutes. At the beginning of the flare, a set of flare loops appear in both the 304 \AA\ and 94 \AA\ channels. In particular, one can notice two bright loops that are located closely in space (Fig.~\ref{fig1}(b)-(c)). The left (southeastern) loop reaches its peak emission about two minutes earlier than the right (northwestern) one (Fig.~\ref{fig1}(a)). After the flare peak, the loops gradually cool down. Here, we focus on the brightest part of the left loop, which produces the first peak of EUV emission and is likely the site of flare energy release (magnetic reconnection).

On the other hand, the magnetic field around the flare region (Fig.~\ref{fig1}(d)) is very complicated, showing adjacent opposite polarities and parasitic polarities. It is known that such a complicated magnetic structure is a favorable place for flare occurrence.

The flare seen in H$\alpha$ and \ion{Ca}{2} 8542 \r{A} brightens nearly simultaneously and cospatially with that in 304 \AA\ and 94 \AA\ during the flare peak time (Fig.~\ref{fig1}(e)-(f)). However, the H$\alpha$ and \ion{Ca}{2} 8542 \r{A} emission is relatively restricted to particular sites where the chromosphere (formation layer of the two lines) gets significantly heated. It is interesting that in the \ion{He}{1} 10830 \r{A} images, there appears a dark thread-like structure in addition to some brightenings seen below the threads (Fig.~\ref{fig1}(g)). The dark threads seem to have a twist and experience an untwisting with the flare development.

\section{Data Analysis and Results}
\subsection{Spectral Analysis of the Flare Region}
In the flaring region, the chromospheric H$\alpha$ and \ion{Ca}{2} 8542 \r{A} lines show obvious asymmetries (see Figs.~\ref{fig3} and \ref{fig4}). In particular, the H$\alpha$ line profile exhibits broadened wings that are higher than the continuum level, and the \ion{Ca}{2} 8542 \r{A} line profile shows a hump in either the red or the blue wing. For these asymmetric line profiles showing a net emission, their contrast profiles, the profiles on the flare subtracted by the pre-flare ones, can be used to study the change in profiles, which is caused by the flare \citep{1987canfield}. Applying the radiative transfer equation with a constant source function $S$ for the flare yields:
\begin{equation}
I_{f}=I_{0}\exp(-\tau_{f})+S[1-\exp(-\tau_{f})],
\end{equation}
where $I_{f}$ is the emergent intensity from the flare, $I_{0}$ is the background intensity from the photosphere, and $\tau_{f}$ is the optical depth of the flare-perturbed atmosphere. At the far wings, the chromosphere is normally transparent, so that the background intensity can roughly be replaced by the pre-flare intensity $I_{n}$, and $\tau_{f}$ is sufficiently less than unity \citep{1984ichimoto}. Then the equation above can be simplified as
\begin{equation}
I_{f}-I_{n}\approx(S-I_{n})\tau_{f}.\label{eq2}
\end{equation}
Therefore, such contrast profiles, in particular at the far wings, can be regarded as the net emission from the flaring plasma at a height where the magnetic energy is released, namely the reconnection site.

We first apply the bisector method to derive the mass flow velocities from the Doppler shifts of the H$\alpha$ line. The bisector method, also referred to as lambdameter, is a simple but often used method to derive Doppler velocities from optically thick line profiles \citep{2013chaeb}. This method makes a horizontal cut at a certain emission intensity on the profile $C(\lambda)$ so that
\begin{equation}
C\left(\lambda_{m}-\frac{\delta\lambda_{b}}{2}\right)=C\left(\lambda_{m}+\frac{\delta\lambda_{b}}{2}\right),
\end{equation}
where $\delta\lambda_{b}$ is the full width of the line cut, and $\lambda_{m}$ is considered as the observed line center. Then the Doppler velocities are derived from the shifts of the line center relative to the theoretical (reference) one. Since the bisector is usually not a straight line, different values of velocities can be obtained for horizontal cuts at different intensity levels \citep{2014hong}, i.e. from line center to wings, which are supposed to represent different heights of the solar atmosphere. Choosing the level for the horizontal cut should balance two requirements as below. On one hand, one needs to choose as far as possible the line wings so that Equation~(\ref{eq2}) can still hold. On the other hand, the noise becomes large at the far wings that can yield a large uncertainty in the derived velocities. For our event, we choose the horizontal cut at line wings with a net emission intensity of 0.2 times the maximum intensity to avoid the invalidity of the method towards the line center as well as the noise at far wings \citep{1995ding}. Such an intensity level corresponds to the chromospheric level where microflares likely occur \citep{2006fang}. Note that our first purpose is to provide a velocity map through a preliminary analysis of the line profiles. Changing the level of the horizontal cut a little can only change slightly the absolute value but not the direction of the derived velocity.

The velocity map at the flare peak time (Fig.~\ref{fig1}(e)) reveals clearly both the downflows and upflows in the flare region, which are located very close to the bright regions seen in H$\alpha$ and \ion{Ca}{2} 8542 \r{A}, also the brightest part of the left flare loop seen in 304 \AA\ and 94 \AA\ (Fig.~\ref{fig1}(b)-(c)). Note that we actually get a time series of velocity maps during the flare, which show that such flows appear impulsively in the flare rise phase and become relatively steady for about two minutes. We select a region of interest (ROI) containing one brightest part with striking opposite velocity signs nearby for further study (the yellow square in Fig.~\ref{fig1}(e)). The size of the ROI is $4\arcsec\times5\arcsec$.

\begin{figure}
    \centering
    \plotone{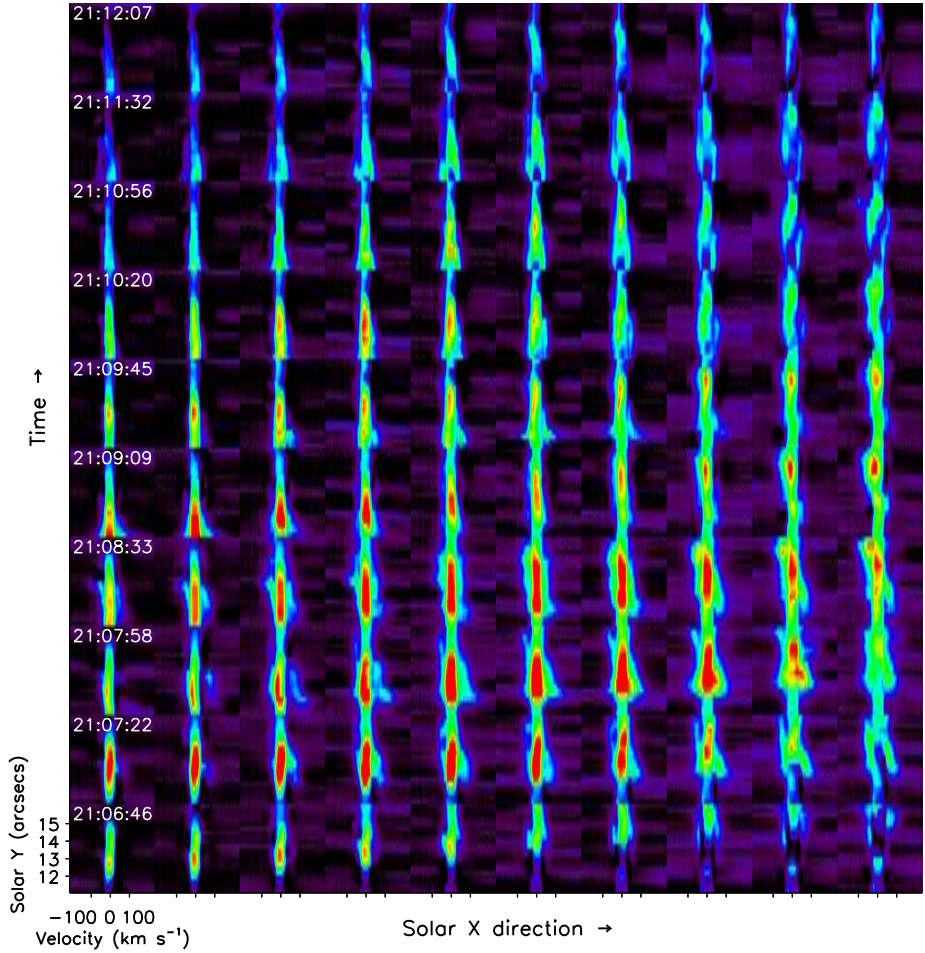}
    \caption{Time sequence of H$\alpha$ net emission spectra. Each sub-panel shows the spectra along a Y-directed line at a specific X-position in the ROI. Each row shows the spectra varying with the X-positions for the same time. Each column shows the spectra varying with time (from bottom to top) for the same X-position. Redshifts are to the right, and blueshifts are to the left. Colours represent different intensity levels.}\label{fig2}
\end{figure}

We divide the ROI into $10\times20$ grids and show the H$\alpha$ contrast profiles at each grid point in a time sequence (Fig.~\ref{fig2}). Line asymmetries are clearly seen during the flare peak time. For example, in a number of subpanels (columns 6--8, row 8), blueshifted emission is seen in the top region while redshifted emission appears in the bottom region. Though the redshifted emission seems stronger than the blueshifted emission, the speed derived from them are similar. The red and blue asymmetries appear nearly simultaneously and at locations only separated by about $3\arcsec$, implying that they are physically related with each other.

For a more quantitative study, we focus on some typical H$\alpha$ line spectra with significant asymmetries along both the scan and slit directions (Fig.~\ref{fig3}(a)), and plot the contrast profiles at two specific positions during the flare peak time (Fig.~\ref{fig3}(c) and (f)). It is interesting that these asymmetric profiles consist of two components, a main (static) component and a second (redshifted or blueshifted) one. Based on the shapes of the line profiles observed in our event, we use a multi-component function, which comprises two Gaussian-shaped components plus a constant background, to fit the contrast profile:
\begin{equation}
C(\lambda)=A_{0}+A_{1}\textrm{exp}\left[-\left(\frac{\lambda-\lambda_{1}}{\sigma_{1}}\right)^{2}\right]
+A_{2}\textrm{exp}\left[-\left(\frac{\lambda-\lambda_{2}}{\sigma_{2}}\right)^{2}\right],
\end{equation}
where $A_{0}$ is the background, $A_{1}$ and $A_{2}$ are the strengths, and $\sigma_{1}$ and $\sigma_{2}$ are the widths of the two components, respectively. In particular, $\lambda_{1}$ and $\lambda_{2}$ are the fitted line centers of the two components.

The procedure `mpfitfun.pro' in the Solar SoftWare package, which can fit a user-supplied model to data using the Levenberg-Marquadt algorithm to solve the least-squares problem, is used for the double Gaussian fitting. The results show Doppler velocities of $\pm$(70--80) km s$^{-1}$ from the second components. The related downflows and upflows should be located in the chromosphere where the H$\alpha$ line is formed \citep{2012leenaarts}. Such a speed is comparable with the local Alfv\'{e}n speed in the chromosphere \citep{2008nishizuka}. Note that the \ion{Ca}{2} 8542 \r{A} profiles show similar asymmetries but with somewhat smaller velocities ($\pm$(40--50) km s$^{-1}$) from the second components (Fig.~\ref{fig4}). These significant bidirectional flows in the flare region are most likely the outflows of magnetic reconnection.

\subsection{The Cause of Line Asymmetries}
\begin{figure}
    \centering
    \plotone{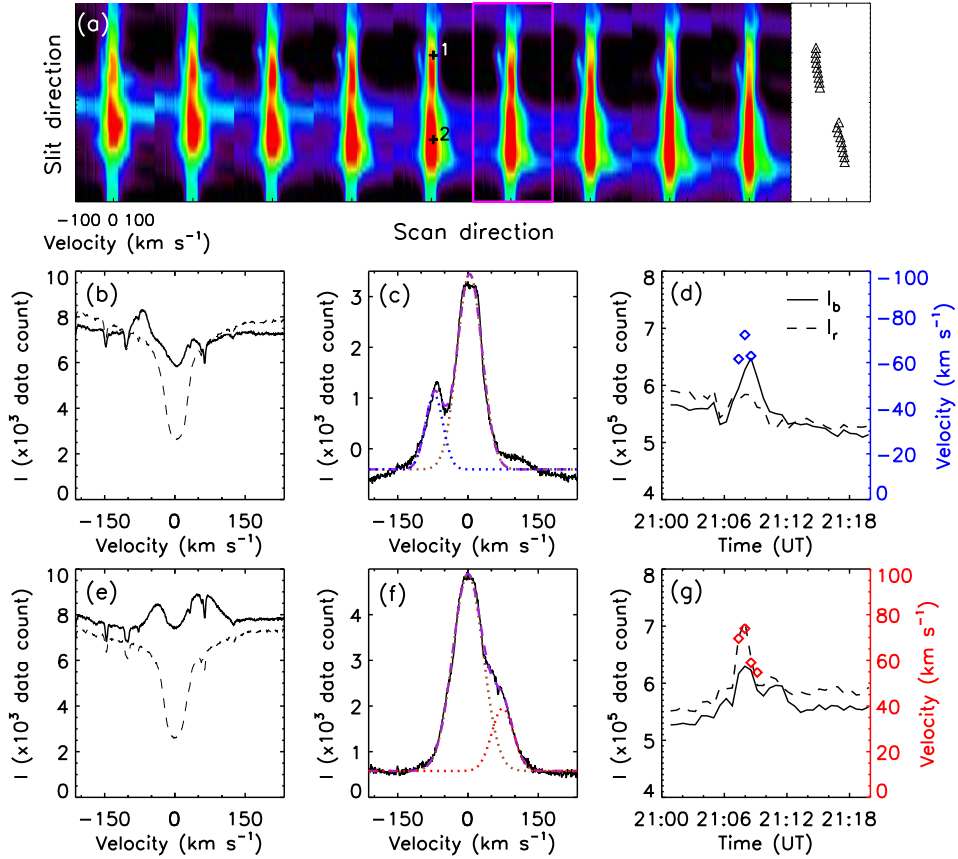}
    \caption{A sample of H$\alpha$ contrast profiles at the flare peak time (21:07:58 UT). (a) Selected H$\alpha$ net emission spectra. Also shown in the right sub-panel is the line-of-sight velocity distribution along the slit derived from a certain position (marked with a pink box). Each sub-panel corresponds to a spatial extension of $6\arcsec.4$ along the slit, and the spacing between two successive slit positions is $0\arcsec.16$. Colors represent different intensity levels. (b) Original line profiles at the flare peak time (solid) and before the flare (dashed) at pixel 1 marked in panel (a). (c) Contrast profile (black solid line) and their two-component fitting (purple dashed line). The main (static) component is shown in brown dotted line, while the blueshifted/redshifted components are shown in blue/red dotted lines, respectively. (d) Time evolution of intensities at the red wing (dashed) and the blue wing (solid), with the integration range of 1 to 2.5 \r{A} and -2.5 to -1 \r{A} from the line center, respectively. The velocities derived from two-component fittings are shown in diamonds. (e)--(g) The same as (b)--(d) but for the profile at pixel 2.}\label{fig3}
\end{figure}

\begin{figure}
    \centering
    \plotone{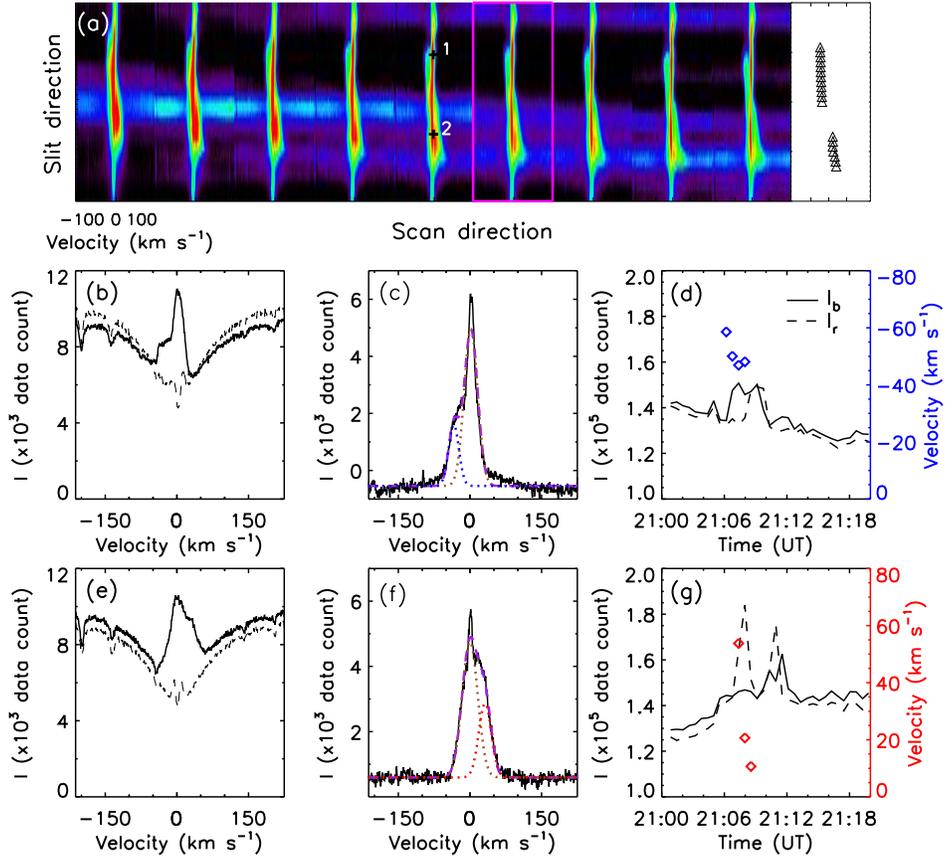}
    \caption{A sample of \ion{Ca}{2} 8542 \r{A} contrast profiles at the flare peak time (21:07:58 UT). All the notations are the same as in Fig.~\ref{fig3}. The red and blue line wing intensities in panels (d) and (g) are integrated from 1 to 1.5 \r{A} and -1.5 to -1 \r{A} from the line center respectively.}\label{fig4}
\end{figure}

Usually, it is difficult to interpret the asymmetries of optically thick lines, in particular the H$\alpha$ line. For the blue asymmetry (larger blue wing emission), it can either be caused by an upward motion of a heated plasma (in emission) or a downward motion of a relatively cool plasma (in absorption). Likewise, the red asymmetry (larger red wing emission) can be produced by either a downward moving plasma in emission or an upward moving plasma in absorption. Such a scenario has been clearly verified by non-LTE calculations \citep{1994heinzel} and more recently by radiative hydrodynamic simulations of the flare atmosphere \citep{2015kuridze}. However, we tend to verify from multi-aspects that in our event, the blue and red asymmetries are related to upflows and downflows of the heated plasma in emission, respectively. First, as can be seen from the asymmetric profiles, there appears a big hump in the blue wing (Fig.~\ref{fig3}(c) and Fig.~\ref{fig4}(c)) or in the red wing (Fig.~\ref{fig3}(f) and Fig.~\ref{fig4}(f)). This is a strong signature of existence of a Doppler-shifted emission component superposed on a static profile at the blue or red wings, while it can hardly be produced by an absorption at the opposite wings. Second, the time evolutions of the line wing intensities hint an emission cause of the line asymmetries. For the red asymmetric profiles, the intensity of the red wing rises more sharply than the blue wing when the flare begins (Fig.~\ref{fig3}(d) and Fig.~\ref{fig4}(d)). This is contrary to the line asymmetries of an absorption cause, in which one might expect a temporary decrease of the intensities at the opposite wings if the increase due to flare heating and the drop due to absorption are not exactly coincident. Third, previous model calculations showed that the line asymmetries may change their sign with the development of the flare owing to the different heating status of the moving plasma \citep{1994heinzel,2015kuridze}. In our event, however, the line asymmetries do not change their sign during the whole flare. Fourth, the same asymmetries appear not only in the H$\alpha$ line but also in the \ion{Ca}{2} 8542 \r{A} line. The latter is formed somewhat lower than the former. To our knowledge, there are no model calculations up to now showing that the blue/red asymmetries of the \ion{Ca}{2} 8542 \r{A} line can be due to downward/upward moving cool plasma in absorption, though theoretically also possible. Finally, in order to further check the possibility of an absorption cause of the line asymmetries, we also do a test to fit the line profiles with a static component plus a negative shifted component (due to absorption) at the opposite wings. This fitting fails to yield reasonable results with the fitting errors much larger than the case of a static component plus a positive shifted component as done in Section~3.1. Considering the above facts, we think that the blue/red line asymmetries in this flare are caused by upward/downward motions of heated mass, which are likely related to reconnection outflows, as will be discussed below.

As for the cause of the mass flows in a flare, one would naturally think of the possibility of chromospheric evaporation and condensation. However, we can exclude this possibility for the following reasons. First, the line-of-sight velocity along a slit, derived from the sequence of H$\alpha$ and \ion{Ca}{2} 8542 \r{A} contrast profiles, shows a gradual (nearly linear) variation in magnitude (Fig.~\ref{fig3}(a) and Fig.~\ref{fig4}(a)). Such a spatial distribution can hardly be explained by the spatially different evaporations, but is more likely a result of plasma acceleration. Second, previous observations show that chromospheric condensation, thought to be responsible for the red asymmetries in H$\alpha$ profiles, rapidly decreases before the H$\alpha$ intensity reaches its maximum \citep{1984ichimoto}. But in our event, the H$\alpha$ red asymmetries seem to exist longer. Third, the most recent radiative hydrodynamic simulations of solar flares show that the H$\alpha$ and \ion{Ca}{2} 8542 \r{A} lines only exhibit red asymmetries in response to chromospheric condensation in the early heating phase, but that they hardly exhibit blue asymmetries responding to chromospheric evaporation containing plasma of very high temperatures \citep{2015rubiodacosta}. This is apparently different from what are revealed in our observations, where the red and blue asymmetries appear almost simultaneously in both chromospheric lines.

\subsection{Magnetic Field Extrapolation}
The three-dimensional magnetic field structure of this flare region is constructed using nonlinear force-free field extrapolations with the optimization method \citep{2000wheatland,2004wiegelmann}. The vector magnetic field on the bottom boundary comes from the Space-weather HMI Active Region Patches (SHARP, \citealt{2014bobra}), in which the $180^{\circ}$ ambiguity of transverse components has already been removed using the minimum energy method \citep{2009leka} and the projection effect has been corrected \citep{1990gary}. The data have been remapped to a heliographic cylindrical equal-area coordinate system. An additional preprocessing is applied for the vector field to meet the force-free and torque-free conditions \citep{2006wiegelmann}. The extrapolation box consists of $230\times200\times200$ uniform grid points in a box, corresponding to a domain of about $175\times152\times152$ Mm$^{3}$.

\begin{figure}
    \centering
    \plotone{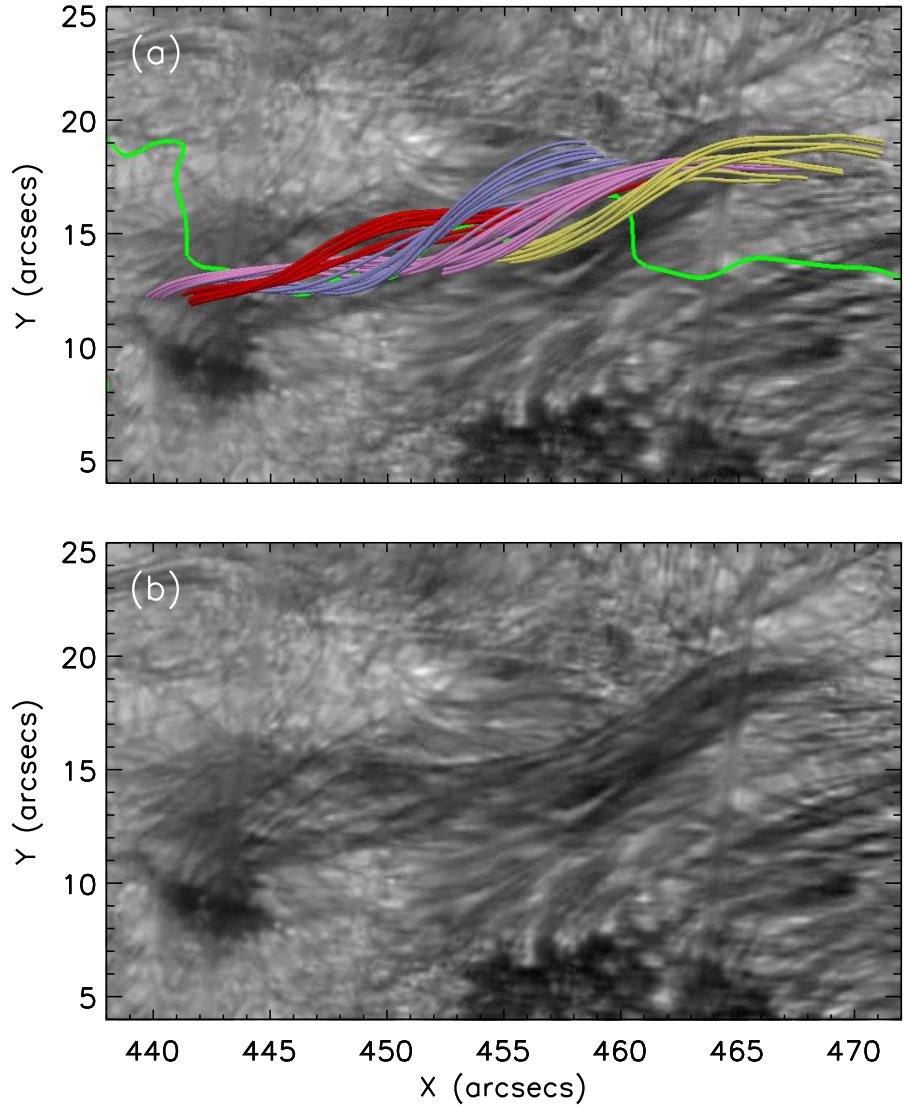}
    \caption{Three-dimensional magnetic structure in the flare region at 20:58:22 UT before the flare. (a) Top view of the magnetic flux rope. Green lines indicate the PIL. (b) The dark threaded structure in the \ion{He}{1} 10830 \r{A} image.}\label{fig5}
\end{figure}

The extrapolation result shows that, before the flare onset, there exist twisted magnetic field lines above the parasitic polarities, which lie along the polarity inversion line (PIL) and overlap well with the dark thread-like structure seen in the \ion{He}{1} 10830 \r{A} images (Fig.~\ref{fig5}). This clearly indicates the existence of a flux rope in the lower solar atmosphere \citep{2015wang}. The height of the flux rope is about 800--1000 km above the solar surface, as judged from both the magnetic structure and the \ion{He}{1} 10830 \r{A} dark threads. Such a low-lying flux rope can serve as a cause of the onset of the reconnection in the chromosphere by its instabilities \citep{2005torok,2010aulanier,2006kliem}. Note that the flare is a confined one since the flux rope failed to erupt finally. And we did not see magnetic flux emergence in the flare region under present resolution.

\section{Summary and Discussions}
We study a microflare based on both high-resolution spectroscopic and imaging observations, focusing on the observational evidence of magnetic reconnection in the chromosphere. The H$\alpha$ contrast profiles show obvious excess emission in the red or blue wings, with a Doppler velocity of $\pm$(70--80) km s$^{-1}$. The \ion{Ca}{2} 8542 \r{A} contrast profiles show similar asymmetric patterns with, however, a slightly lower velocity. Such a difference could originate from the different formation heights of the two lines \citep{2012leenaarts,2008cauzzi}. Nevertheless, the qualitatively similar behaviours shown in the two chromospheric lines do suggest that the mass flows lie in the chromosphere. The co-existence of both upflows and downflows, located closely in space and with speeds comparable to the local Alfv\'{e}n speed, indicates that they are a pair of outflows resulting from the reconnection in the chromosphere.

Previous observations have yielded spectroscopic measurements of reconnection outflows. \citet{2007wang} showed fast flows with speeds of $\sim$900--3500 km s$^{-1}$ in a flare. Redshifted outflows of about 125 km s$^{-1}$ in a flare were also reported \citep{2014tian}. In addition, \citet{1997innes} showed bidirectional outflows ($\sim$100 km s$^{-1}$) in transition region explosive events. In our event, we find bidirectional outflows with a Doppler velocity of $\pm$(70--80) km s$^{-1}$ in the H$\alpha$ line. From the magnetic field extrapolation, the flux rope is found to lie at a height of $\sim$900 km, where the ambient magnetic field is $\sim$600 G and the mass density is $2.34\times10^{-10}$ g cm$^{-3}$ if we take the F1 model of a weak flare \citep{1980machado} as a reference model. Then, the local Alfv\'{e}n speed at the reconnection site, if just below the flux rope, is calculated to be $\sim$110 km s$^{-1}$. The speeds of the bidirectional outflows are then roughly comparable with the local Alfv\'{e}n speed, indicating that the outflows are most likely driven by the flare-related magnetic reconnection.

Therefore, our results show clear evidence of magnetic reconnection in the chromosphere leading to the occurrence of a microflare. The chain of evidence includes the trigger of reconnection, a twisted flux rope above the flaring site, and the effect of reconnection, bidirectional outflows. In fact, magnetic reconnection in the lower atmosphere has been suggested to be the cause of some small-scale activities like Ellerman bombs \citep{2009archontis,2015vissers} and hot explosions \citep{2014peter}. Our observations suggest that flare-related magnetic reconnection can also proceed in the chromosphere. Moreover, we confirm that the basic mechanisms for a microflare are similar to those for a major flare occurring mostly in the corona, regardless of the quite different energies between them.

\acknowledgments
We are very grateful to the referee for valuable comments that helped improve the paper. The observation program was supported by the Strategic Priority Research Program --- The Emergence of Cosmological Structures of the Chinese Academy of Sciences, Grant No. XDB09000000. The authors thank the BBSO staff for their help during the observations. SDO is a mission of NASA's Living With a Star Program. This work was also supported by NSFC under grants 11303016, 11373023, 11403011 and 11533005, and NKBRSF under grants 2011CB811402 and 2014CB744203. W.~C. acknowledges the support of the US NSF (AGS-0847126) and NASA (NNX13AG14G). J.~H. would also like to thank Donguk~Song for his help in FISS data analysis.

\clearpage

\begin{thebibliography}{dummy}
\bibitem[Acton et al.(1992)]{1992acton} Acton, L.~W., Feldman,
U., Bruner, M.~E., et al.\ 1992, \pasj, 44, L71


\bibitem[Archontis
\& Hood(2009)]{2009archontis} Archontis, V., \& Hood, A.~W.\ 2009, \aap, 508, 1469


\bibitem[Aulanier et al.(2010)]{2010aulanier} Aulanier, G.,
T{\"o}r{\"o}k, T., D{\'e}moulin, P., \& DeLuca, E.~E.\ 2010, \apj, 708, 314


\bibitem[Bentley et al.(1994)]{1994bentley} Bentley, R.~D.,
Doschek, G.~A., Simnett, G.~M., et al.\ 1994, \apjl, 421, L55


\bibitem[Bobra et al.(2014)]{2014bobra} Bobra, M.~G., Sun, X.,
Hoeksema, J.~T., et al.\ 2014, \solphys, 289, 3549

\bibitem[Canfield
\& Metcalf(1987)]{1987canfield} Canfield, R.~C., \& Metcalf, T.~R.\ 1987, \apj, 321, 586

\bibitem[Cao et al.(2012)]{2012cao} Cao, W., Goode, P.~R., Ahn,
K., et al.\ 2012, Second ATST-EAST Meeting: Magnetic Fields from the
Photosphere to the Corona., 463, 291


\bibitem[Cao et al.(2010)]{2010cao} Cao, W., Gorceix, N.,
Coulter, R., et al.\ 2010, Astronomische Nachrichten, 331, 636


\bibitem[Carmichael(1964)]{1964carmichael} Carmichael, H.\ 1964, NASA
Special Publication, 50, 451


\bibitem[Cauzzi et
al.(2008)]{2008cauzzi} Cauzzi, G., Reardon, K.~P., Uitenbroek, H., et al.\ 2008, \aap, 480, 515


\bibitem[Chae et al.(2013a)]{2013chae} Chae, J., Park, H.-M.,
Ahn, K., et al.\ 2013, \solphys, 288, 1

\bibitem[Chae et al.(2013b)]{2013chaeb} Chae, J., Park, H.-M.,
Ahn, K., et al.\ 2013, \solphys, 288, 89

\bibitem[Cheng et al.(2011)]{2011cheng} Cheng, X., Zhang, J.,
Liu, Y., \& Ding, M.~D.\ 2011, \apjl, 732, L25


\bibitem[Ding et al.(1995)]{1995ding} Ding, M.~D., Fang, C.,
\& Huang, Y.~R.\ 1995, \solphys, 158, 81

\bibitem[Doschek et al.(1995)]{1995doschek} Doschek, G.~A., Strong,
K.~T., \& Tsuneta, S.\ 1995, \apj, 440, 370


\bibitem[Fang et al.(2006a)]{2006fanga} Fang, C., Tang, Y.~H., Xu,
Z., Ding, M.~D., \& Chen, P.~F.\ 2006, \apj, 643, 1325


\bibitem[Fang et al.(2006b)]{2006fang} Fang, C., Tang, Y.-H.,
\& Xu, Z.\ 2006, \cjaa, 6, 597

\bibitem[Gary
\& Hagyard(1990)]{1990gary} Gary, G.~A., \& Hagyard, M.~J.\ 1990, \solphys, 126, 21


\bibitem[Goode
\& Cao(2012)]{2012goode} Goode, P.~R., \& Cao, W.\ 2012, \procspie, 8444, 844403

\bibitem[Heinzel et al.(1994)]{1994heinzel} Heinzel, P., Karlicky,
M., Kotrc, P., \& Svestka, Z.\ 1994, \solphys, 152, 393

\bibitem[Hirayama(1974)]{1974hirayama} Hirayama, T.\ 1974, \solphys,
34, 323


\bibitem[Hong et al.(2014)]{2014hong} Hong, J., Ding, M.~D., Li,
Y., Fang, C., \& Cao, W.\ 2014, \apj, 792, 13


\bibitem[Ichimoto
\& Kurokawa(1984)]{1984ichimoto} Ichimoto, K., \& Kurokawa, H.\ 1984, \solphys, 93, 105


\bibitem[Innes et al.(1997)]{1997innes} Innes, D.~E., Inhester,
B., Axford, W.~I., \& Wilhelm, K.\ 1997, \nat, 386, 811


\bibitem[Kliem
\& T{\"o}r{\"o}k(2006)]{2006kliem} Kliem, B., T{\"o}r{\"o}k, T.\ 2006, Physical Review Letters, 96, 255002


\bibitem[Kopp
\& Pneuman(1976)]{1976kopp} Kopp, R.~A., \& Pneuman, G.~W.\ 1976, \solphys, 50, 85


\bibitem[Kuridze et al.(2015)]{2015kuridze} Kuridze, D.,
Mathioudakis, M., Sim{\~o}es, P.~J.~A., et al.\ 2015, \apj, 813, 125

\bibitem[Leenaarts et al.(2012)]{2012leenaarts} Leenaarts, J.,
Carlsson, M., \& Rouppe van der Voort, L.\ 2012, \apj, 749, 136


\bibitem[Leka et al.(2009)]{2009leka} Leka, K.~D., Barnes, G.,
Crouch, A.~D., et al.\ 2009, \solphys, 260, 83


\bibitem[Lemen et al.(2012)]{2012lemen} Lemen, J.~R., Title,
A.~M., Akin, D.~J., et al.\ 2012, \solphys, 275, 17


\bibitem[Liu et al.(2013)]{2013liu} Liu, W., Chen, Q.,
\& Petrosian, V.\ 2013, \apj, 767, 168

\bibitem[Machado et al.(1980)]{1980machado} Machado, M.~E., Avrett,
E.~H., Vernazza, J.~E., \& Noyes, R.~W.\ 1980, \apj, 242, 336

\bibitem[Masuda et al.(1994)]{1994masuda} Masuda, S., Kosugi, T.,
Hara, H., Tsuneta, S., \& Ogawara, Y.\ 1994, \nat, 371, 495


\bibitem[Nishizuka et al.(2008)]{2008nishizuka} Nishizuka, N.,
Shimizu, M., Nakamura, T., et al.\ 2008, \apjl, 683, L83


\bibitem[Pesnell et al.(2012)]{2012pesnell} Pesnell, W.~D.,
Thompson, B.~J., \& Chamberlin, P.~C.\ 2012, \solphys, 275, 3


\bibitem[Peter et al.(2014)]{2014peter} Peter, H., Tian, H.,
Curdt, W., et al.\ 2014, Science, 346, 1255726


\bibitem[Reeves et al.(2015)]{2015reeves} Reeves, K.~K., McCauley,
P.~I., \& Tian, H.\ 2015, \apj, 807, 7


\bibitem[Rubio da Costa et al.(2015)]{2015rubiodacosta} Rubio da Costa,
F., Kleint, L., Petrosian, V., Sainz Dalda, A.,
\& Liu, W.\ 2015, \apj, 804, 56


\bibitem[Sakao et al.(1992)]{1992sakao} Sakao, T., Kosugi, T.,
Masuda, S., et al.\ 1992, \pasj, 44, L83


\bibitem[Schou et al.(2012)]{2012schou} Schou, J., Scherrer,
P.~H., Bush, R.~I., et al.\ 2012, \solphys, 275, 229


\bibitem[Shibata(1999)]{1999shibata} Shibata, K.\ 1999, \apss, 264, 129


\bibitem[Sturrock(1999)]{1999sturrock} Sturrock, P.~A.\ 1999, \apj,
521, 451


\bibitem[Sturrock(1966)]{1966sturrock} Sturrock, P.~A.\ 1966, \nat,
211, 695


\bibitem[Sun et al.(2015)]{2015sun} Sun, J.~Q., Cheng, X.,
Ding, M.~D., et al.\ 2015, Nature Communications, 6, 7598


\bibitem[Tandberg-Hanssen
\& Emslie(1988)]{1988tandberg} Tandberg-Hanssen, E., \& Emslie, A.~G.\ 1988, Cambridge and New York, Cambridge University Press, 1988


\bibitem[Tian et al.(2014)]{2014tian} Tian, H., Li, G., Reeves,
K.~K., et al.\ 2014, \apjl, 797, L14

\bibitem[T{\"o}r{\"o}k
\& Kliem(2005)]{2005torok} T{\"o}r{\"o}k, T., \& Kliem, B.\ 2005, \apjl, 630, L97


\bibitem[Tsuneta et al.(1992)]{1992tsuneta} Tsuneta, S., Hara, H.,
Shimizu, T., et al.\ 1992, \pasj, 44, L63


\bibitem[Vissers et al.(2015)]{2015vissers} Vissers, G.~J.~M.,
Rouppe van der Voort, L.~H.~M., Rutten, R.~J., Carlsson, M.,
\& De Pontieu, B.\ 2015, \apj, 812, 11

\bibitem[Wang et al.(2015)]{2015wang} Wang, H., Cao, W.,
Liu, C., et al.\ 2015, Nature Communications, 7, 7008


\bibitem[Wang et al.(2007)]{2007wang} Wang, T., Sui, L.,
\& Qiu, J.\ 2007, \apjl, 661, L207

\bibitem[Wheatland et al.(2000)]{2000wheatland} Wheatland, M.~S.,
Sturrock, P.~A., \& Roumeliotis, G.\ 2000, \apj, 540, 1150


\bibitem[Wiegelmann(2004)]{2004wiegelmann} Wiegelmann, T.\ 2004,
\solphys, 219, 87


\bibitem[Wiegelmann et al.(2006)]{2006wiegelmann} Wiegelmann, T.,
Inhester, B., \& Sakurai, T.\ 2006, \solphys, 233, 215



\bibitem[Yokoyama et al.(2001)]{2001yokoyama} Yokoyama, T., Akita,
K., Morimoto, T., Inoue, K., \& Newmark, J.\ 2001, \apjl, 546, L69

\end{thebibliography}
\end{document}